\documentclass[sigconf]{acmart}

\AtBeginDocument{%
  \providecommand\BibTeX{{%
    \normalfont B\kern-0.5em{\scshape i\kern-0.25em b}\kern-0.8em\TeX}}}

\setcopyright{acmcopyright}
\copyrightyear{2023}
\acmYear{2023}
\acmDOI{XXXXXXX.XXXXXXX}

\acmConference[Conference acronym 'XX]{Make sure to enter the correct
  conference title from your rights confirmation emai}{June 03--05,
  2018}{Woodstock, NY}
\acmPrice{15.00}
\acmISBN{978-1-4503-XXXX-X/18/06}

\usepackage{amsmath,amsthm}
\usepackage{mathtools}
\newcommand{\dd}{\mathrm{d}}

\theoremstyle{plain}

\theoremstyle{definition}

\theoremstyle{remark}

\begin{document}
%
\title[Interpretable (not just posthoc-explainable) readmissions model for discharge placement]{Interpretable (not just posthoc-explainable) medical claims modeling for discharge placement to reduce preventable all-cause readmissions or death}

\author{Joshua C. Chang}
\email{josh@mederrata.com}
\orcid{0000-0001-9690-9179}
\author{Ted L. Chang}
\email{ted@mederrata.com}
\affiliation{%
  \institution{Mederrata Research and Sound Prediction}
  \country{USA}
}

\author{Carson C. Chow}
\email{carson.chow@nih.gov}
\orcid{0000-0003-1463-9553}
\affiliation{%
  \institution{NIH NIDDK and Mederrata Research}
  \country{USA}
}

\author{Rohit Mahajan}
\email{ro@mederrata.com}
\author{Sonya Mahajan}
\email{sonya@mederrata.com}
\affiliation{%
  \institution{Mederrata Research and Sound Prediction}
  \country{USA}
}

\author{Joe Maisog}
\email{bravas02@gmail.com}
\affiliation{%
  \institution{Blue Health Intelligence}
  \country{USA}
}

\author{Shashaank Vattikuti}
\email{shashaank76@gmail.com}
\affiliation{%
  \institution{Walter Reed Army Institute of Research}
  \country{USA}
}

\author{Hongjing Xia}
\email{hongjing@mederrata.com}
\affiliation{%
  \institution{Sound Prediction and Mederrata Research}
  \country{USA}
}

\renewcommand{\shortauthors}{Chang, Chang, Chow, Mahajan, Mahajan, Maisog, Vattikuti, Xia}

\newcommand{\bu}{\mathbf{u}}
\newcommand{\bv}{\mathbf{v}}
\newcommand{\bb}{\mathbf{b}}

\newcommand{\bz}{\mathbf{z}}
\newcommand{\bx}{\mathbf{x}}
\newcommand{\bw}{\mathbf{w}}
\newcommand{\bmu}{\boldsymbol{\mu}}
\newcommand{\bV}{\mathbf{V}}

\newcommand{\bZ}{\mathbf{Z}}
\newcommand{\bX}{\mathbf{X}}
\newcommand{\bW}{\mathbf{W}}
\newcommand{\bS}{\mathbf{S}}
\newcommand{\bU}{\mathbf{U}}
\newcommand{\bbeta}{\boldsymbol{\beta}}
\newcommand{\btheta}{\boldsymbol{\theta}}
\newcommand{\bphi}{\boldsymbol{\phi}}
\newcommand{\balpha}{\boldsymbol{\alpha}}
\newcommand{\bM}{\mathbf{M}}
\newcommand{\bI}{\mathbf{I}}
\newcommand{\R}{\mathbb{R}}

\begin{abstract}

We developed an inherently interpretable multilevel Bayesian  framework for representing variation in regression coefficients that mimics the piecewise linearity of ReLU-activated deep neural networks. We used the framework to formulate a survival model for using medical claims to predict hospital readmission and death that focuses on discharge placement, adjusting for confounding in estimating causal local average treatment effects.
We trained the model on a 5\% sample of Medicare beneficiaries from 2008 and 2011, based on their 2009--2011 inpatient episodes, and then tested the model on 2012 episodes.
The model scored an AUROC of approximately 0.76 on predicting  all-cause readmissions -- defined using official Centers for Medicare and Medicaid Services (CMS) methodology -- or death within 30-days of discharge, 
being competitive against XGBoost and a Bayesian deep neural network, demonstrating that one need-not sacrifice interpretability for accuracy.
Crucially, as a regression model, we provide what blackboxes cannot -- the exact gold-standard global interpretation of the model, identifying relative risk factors and quantifying the effect of discharge placement. We also show that the posthoc explainer SHAP fails to provide accurate explanations.

\end{abstract}

\maketitle
\section{Introduction}
\label{introduction}

Preventable readmission after hospital discharge is costly.
In 2011, for adult 30-day all cause hospital readmission in the United States, the cost was about \$41.3 billion~\citep{hinesConditionsLargestNumber2014}. 
To improve outcomes, Medicare, through its Hospital Readmissions Reduction Program~\citep{mcilvennanHospitalReadmissionsReduction2015},
penalizes providers for readmissions that occur within the 30-days after discharge; penalties have spurred interest in interventions surrounding transitions of care including
 discharge planning services such as hand-offs to less-intensive healthcare institutions.
Population-level medical claims data make it possible to assess the efficacy of these interventions retroactively. This manuscript focuses on the problem of deciding discharge placement for individuals in order to prevent readmission or death.

\noindent\textbf{Readmission models:} 
A recent review~\citep{huangApplicationMachineLearning2021} surveyed properties of readmission models in the literature. 
By and large, they found no model type to consistently predict more-accurately than others, though several studies have reported marginal improvements using either XGBoost or neural networks over logistic regression
~\citep{shameerPredictiveModelingHospital2016,jameiPredictingAllcauseRisk2017, allamNeuralNetworksLogistic2019,liuPredicting30dayHospital2020,minPredictiveModelingHospital2019,futomaComparisonModelsPredicting2015,larssonAdvancedMachineLearner2021}. 
Generally, the literature has focused on 30-day readmissions, though nuances in how readmission is defined complicate direct performance comparisons. 
Models based on medical claims data typically achieved area under the receiver operator characteristic (AUROC) of approximately 0.7 for predicting their version of all cause 30-day readmission.

Another factor that complicates the direct comparison of modeling efforts is differences in datasets -- and hence the underlying patient populations and predictors.
We are aware of two readmission studies performed on datasets identical to ours.
\citet{lahlouExplainableHealthRisk2021} created an attention-based neural network for predicting admissions after discharge within 30-days and reported an AUROC value of 0.81, however, they did not distinguish between transfers, planned admissions, and acute admissions in their outcome label so they solve a different problem that is of less practical utility. More related to our work, \citet{mackayApplicationMachineLearning2021} developed XGBoost models for predicting a set of adverse events, reporting an AUROC of 0.73 for all-cause readmission prediction.

Yet, having a high AUROC is insufficient for making a model useful.
Prerequisites for utility include the ability to understand predictions, assess validity, and derive actions.
Model interpretability is a means to these ends.
Most studies surveyed were aware of the importance of model interpretability, regardless of whether they produced interpretable models. 
Studies that claim interpretability for their blackbox solutions only offer ``posthoc explainability,'' a catch-all phrase for narratives generated in order to promote a sense that a model is interpretable when it is not.

\begin{figure*}
    \centering
    \includegraphics[width=0.95\textwidth]{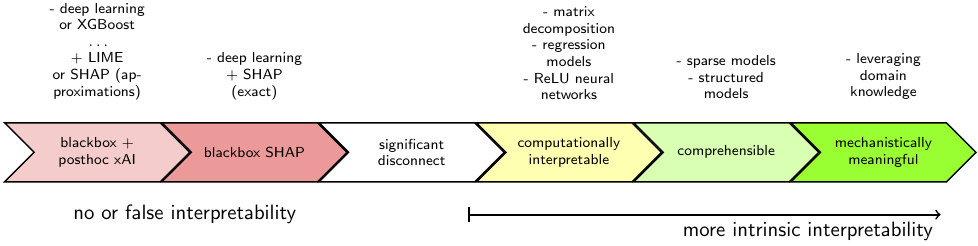}
    \caption{\textbf{Model interpretability} lies along a spectrum with a clear chasm existing between intrinsically interpretable models others.  Models without intrinsic interpretability rely on unreliable approximation techniques for crafting explanations. More-interpretable models are more trustworthy and insightful.}
    \label{fig:interpretability}
\end{figure*}

\noindent\textbf{Interpretability: } The goal of interpretable modeling is to produce predictions that an end-user can understand~\citep{rudinStopExplainingBlack2019,rudinAlgorithmsInterpretableMachine2014}, which is a prerequisite for making a prediction actionable.
One necessary yet insufficient aspect of intrinsic model interpretability is feature attribution.
Blackbox models do not admit feature attributions without the use of unreliable approximations. Conversely, feature attribution is exact in regression models, where each model coefficient has the unequivocal interpretation as the conditional expected change of the response corresponding to a given unit of change in the predictor, while fixing the other predictors.
 For this reason, even ignoring attributes beyond feature attribution, a significant disconnect already separates blackbox models from inherently interpretable models. 
 Fig~\ref{fig:interpretability} is a representation of the spectrum of interpretability focusing on structured data problems in healthcare.

While the definition of interpretability varies according to problem domain, 
all notions of interpretability require a basic ability to parse the computations behind a model's predictions in terms of the input data features.
We refer to this fundamental aspect of interpretability as ``computational interpretability.''
Computational interpretability is a necessary yet insufficient attribute for prediction comprehensibility. ReLU-activated neural networks, matrix composition methods like principle components analysis (PCA), and large multiple regression models are computationally interpretable, whereas Deep Learning (DL) models in-general and ensemble trees methods like XGBoost are not.

However, knowing how a prediction is computed from individual features does not automatically make the prediction comprehensible -- it may still difficult to understand how a model behaves as there is a limit to the capacity of information that humans can process simultaneously~\citep{millerMagicalNumberSeven1956}. 
\citet{sudjiantoDesigningInherentlyInterpretable2021} note that additivity, sparsity, linearity, smoothness, monotonicity, and visualizability are some attributes of interpretable models that are also comprehensible. 
Each of these attributes can be enforced through suitable modeling constraints. 

The highest bar for interpretability is for a model to be mechanistically meaningful. These models often leverage domain knowledge and are capable of providing deep and robust insights. They also often justify causal interpretations~\citep{petersCausalModelsDynamical2020}. Even if one can truly understand a model, one often cannot act on it.
To be directly actionable, a model also needs to adjust for biases in the data so that its prediction of the effects of interventions can be interpreted causally~\citep{pearlCausalInferenceStatistics2009}.
Yet, independent of causal validity, predictive model interpretability is still important because it allows practitioners to better understand the risks and biases of a given model.

\noindent \textbf{Posthoc explainable-AI (xAI):} Posthoc xAI is a set of techniques to market uninterpretable blackbox models as interpretable (Fig.~\ref{fig:interpretability}). 
 The most popular xAI methods (LIME~\citep{ribeiroWhyShouldTrust2016} and SHAP~\citep{lipovetskyAnalysisRegressionGame2001,dattaAlgorithmicTransparencyQuantitative2016}), use approximations~\citep{lundbergUnifiedApproachInterpreting2017,aasExplainingIndividualPredictions2021} to provide narratives of feature importance within a prediction.
 Other methods such as attention~\citep{niuReviewAttentionMechanism2021} build an explanation mechanism as a module within a blackbox model in order to more-easily compute them~\citep{jainAttentionNotExplanation2019,zhouFeatureAttributionMethods2022}.
 Narratives, convincing as they might seem, are not necessarily true. In fact, researchers have shown~\citep{laugelDangersPosthocInterpretability2019,kumarProblemsShapleyvaluebasedExplanations2020,slackFoolingLIMESHAP2020,alvarez-melisRobustnessInterpretabilityMethods2018,zhouFeatureAttributionMethods2022} that these methods provide imprecise and unreliable explanations of models, and often disagree. Aptly, \citet{krishnaDisagreementProblemExplainable2022} coined this the ``disagreement problem" with posthoc interpretability and conducted a survey of real world data scientists finding no consistent or principled method to handle these inconsistencies. As~\citet{rudinStopExplainingBlack2019} notes, ``an explanation model that is correct 90\% of the
time is wrong 10\% of the time.'' Despite marketing claims, xAI does not carry blackbox models across even a very minimal bar of requirements for interpretability.
If an explanation is not true to one's model, any sense that the model is comprehensible is based on faulty information.

\noindent\textbf{Blackbox models:} Methods such as Deep Learning (DL) and ensemble boosted trees (XGBoost, LightGBM, others) can model nonlinearities.
When copious training data is available, these methods yield models that are more expressive than traditional generalized linear models. Most-generally, blackbox models like DL and ensemble trees are nonlinear kernel machines (function interpolations)~\citep{domingosEveryModelLearned2020}.  The convoluted nature of their interpolations makes these models uninterpretable.
 Massive investment exists in these models because of their predictive performance and low effort requirement. This existing investment, the challenge of creating truly interpretable models, and a myth that blackboxes perform better than interpretable models~\citep{rudinStopExplainingBlack2019}, incentivize the marketing of posthoc-xAI as an alternative to interpretable modeling. In finance, a similarly high-stakes domain, there has been wide resistance to blackbox modeling, formalized recently in model risk management guidelines published by  ~\citet{theofficeofthecomptrollerofthecurrencyoccComptrollerHandbookModel2021}. We should also be wary of the use of these models in healthcare, where the risk to patients requires truly trustworthy solutions. 

Blackboxes provide clues on how to extend traditional linear models. DL is the application of artificial neural networks (ANNs) to prediction problems. ANNs consist of sequences (or more generally of graphs) of successive affine matrix arithmetic operations, sandwiched between activation functions. In general, these methods are blackboxes, with the exception of ReLU-activated neural networks (ReLU-nets for short). Examining ReLU-nets elucidates the nature of how DL captures nonlinearities. 
ReLU-nets use the activation function
\begin{equation}
    \textrm{ReLU}(x) = \
    0  \ \textrm{if} \ x\leq 0 \quad \textrm{or} \quad x \ \textrm{otherwise}. 
\end{equation}
In these models, ReLU is independently applied to each matrix coordinate after each successive matrix operation. The output of the function is nonzero if and only if a linear combination of the elements computed by the prior layer are positive. Hence, ReLU defines an inequality over quantities within the model -- applied to each coordinate within each layer, ReLU defines recursive sets of inequalities. These inequalities collectively segment the training data into disjoint regions.
In sum, ReLU-nets are composed of regionally-disjoint generalized linear models -- each of which is interpreted in the same manner as linear regression. Hence, ReLU-nets are computationally interpretable. The salient nonlinearity of these models is locality. To interpret a specific prediction given by these models, one needs to map the input to a particular linear submodel. Then, conditional on this mapping, a ReLU net is locally a simple generalized multiple linear regression model. Observing this fact, \citet{sudjiantoUnwrappingBlackBox2020} provides a tool for exactly interpreting trained ReLU neural networks, by unwrapping the cascades of inequalities. 
In this manuscript we mimic this property of ReLU-nets within a well-controlled multilevel Bayesian regression framework in order to gain expressiveness while prioritizing interpretability.

\section{Methods}

We generalize the classic readmission problem of within 30 days of discharge, to the likelihood of readmission at any arbitrary day after discharge.
To this end, our objective is to characterize the statistics of the inter-inpatient wait time $T_n$.
Additionally, we focus on identifying the effects of discharge placement, representing the choices symbolically as $I_n$, ranked in terms of health acuity:
(0) discharge home, 
    (1) discharge home with home health,
    (2) discharge to skilled nursing,
    (3) intermediate care/critical access,
    (4) long term care,
    (5) other less-acute inpatient.
The issue that complicates the estimation of discharge placement effects is unobserved confounding -- providers use the patient's health status in order to decide  placement. 
To resolve the treatment assignment bias, 
we model the joint outcomes
\begin{align}
T_n \sim f(T_n | \boldsymbol{x}_n, \boldsymbol{I}_n, \boldsymbol\alpha_n, \boldsymbol\beta_n, \boldsymbol\gamma_n) \nonumber \\
I_n \sim g(I_n | \boldsymbol{x}_n, \boldsymbol\nu_n, \boldsymbol\xi_n),
\label{eq:dualoutcome}
\end{align}
where $\boldsymbol{x}_n \in \mathbb{R}^p$ is a covariate vector and we explicitly adjust for assignment bias. 
Note that we distinguish between the scalar-valued $I_n$, which corresponds to the list of interventions above, and the vector valued $\boldsymbol{I}_n$ which we will explain later in this manuscript.
For the sake of interpretability, we formulate $f$ and $g$ in Eq.~\ref{eq:dualoutcome} as hierarchical multilevel Bayesian generalized linear regression models,
However, to increase expressivity by introducing the type of nonlinearity seen in ReLU-nets, we allow all of the model parameters $\boldsymbol{\alpha}_n, \boldsymbol{\beta}_n, \boldsymbol{\gamma}_n, \boldsymbol{\nu}_n, \boldsymbol{\xi}_n$ to vary locally~\citep{hastieVaryingCoefficientModels1993,fanStatisticalMethodsVarying2008,liEstimationVaryingCoefficient2021} across regions defined by $\boldsymbol{x}_n$, in ways that comport with domain knowledge.

\noindent\textbf{Data Preprocessing:}
The available dataset, the CMS Limited Dataset (CMS LDS), consists of a national 5\% beneficiary sample of Medicare FFS Part A and B claims from 2008 to 2012.
The 2008 claims had only quarter date specificity so we used them solely to fill out the medical history for 2009 inpatient stays, by assuming that each 2008 claim fell in the middle of its given quarter. We trained the readmission models on 2009 -- 2011 admissions, and evaluated the models on 2012 admissions.

\begin{figure}[h]
    \centering
    \includegraphics[width=0.85\linewidth]{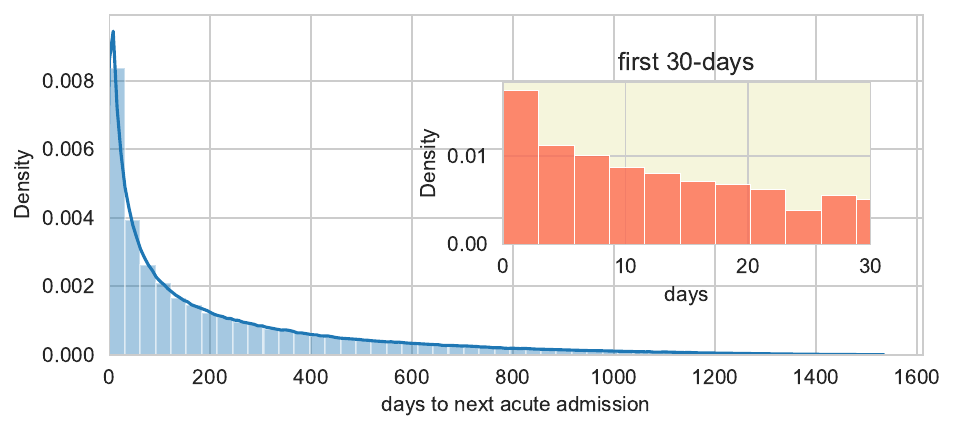}
    \caption{\textbf{Histogram of the wait time after discharge to the next acute admission or death.}}
    \label{fig:histogram}
\end{figure}

After grouping claims into coherent episodes, based on date, provider, and patient overlap, we filtered for inpatient-specific episodes with certain characteristics to use as index admissions.
We retained only episodes where the patient had a continuous prior year of Part A/B enrollment.
We also excluded episodes from consideration as index episodes if they did not correspond to discharges to less-intensive care (excluding death and most inpatient-to-inpatient transfers).
Additionally, we used the official CMS methodology for determining whether each episode is a planned admission, acute admission, or potentially planned admission~\citep{cms2015MeasureInformation2015}.
For each episode we then computed the waiting time to either the next unplanned acute episode or death, or until censorship due to the end of the observation window.
In the end, the training dataset consisted of approximately 1.2 million inpatient episodes, of which approximately 17\% were followed by an unplanned acute inpatient episode or death within 30 days. The histogram of the wait times is presented in Fig.~\ref{fig:histogram}.

For each episode, we collected all billing codes, creating lists of concurrent procedure and diagnostic codes.
Additionally, we collected the preceding four quarters of history for each episode, aggregating billing codes on a lagged quarterly basis.

\noindent\textbf{Feature engineering:}
Medical claims data consists of series of billing codes in several dialects (ICD9/10, HCPCS, RUG, HIPPS, etc). We down-sampled diagnostic (Dx) and procedure (Tx) codes, from their original dialects to multilevel Clinical Classification Software (CCS) codes~\citep{ahrqHCUPUSToolsSoftware2022}. CCS codes are clinically curated hierarchical categories that are  more tractable for analysis and interpretation.
Mapping to CCS drastically removes redundancy in the vocabulary of the dataset and helps to separate the health-specific information in billing codes from noisy reimbursement-specific details.

We used AHRQ Healthcare Cost and Utilization Project (HCUP) databases in order to tag codes for comorbidities, chronic conditions, surgical flags, utilization flags, and procedure flags. Included within skilled nursing facility (SNF) and home health (HH) claim codes are also activities of daily living (ADL) assessments. 
We converted these codes to ADL scores, where higher scores correspond to lower functional ability.
We also incorporated CMS's risk adjustment methodology, hierarchical condition categories (HCC), as model predictors.
The CMS LDS contains beneficiary county codes that we used to incorporate the urban rural index and social economic scale as model features. 
Together with beneficiary race information and Medicaid state buy-in, these variables allowed for some measure of social determinants of health. 

We encoded CCS and other code mappings into numerical vectors by counting the incidences of each permissible code.
In the case of CCS, which is multilevel, we truncated codes at each of the first two levels and counted at each level. 
Altogether, the numerically encoded derived features constituted a vector of size $p=1072,$ which encompassed both concurrent episode codes and the past four quarters of history, where CCS was truncated to the first level for history.

\noindent\textbf{Feature quantization:} To improve model interpretability, we made an effort to place all model parameters (log hazard ratios) on the same scale so that the magnitudes of all regression coefficients are directly comparable.
In examining our derived data features, we found that they were predominantly sparse and heavy tailed.
When fitting a logistic regression model to these data features, the model fit poorly to observations with large counts.
Theses findings, and our desire to optimize model interpretability, led us to quantize all numerical variables so that the input variables into the model are entirely binary.
To this end, we first computed the percentiles for each feature across the entire dataset.
Then we re-coded each quantity into a series of binary variables corresponding to inequalities, where the cutoffs were determined by examining each variables at a set of quantiles and eliminating duplicate values.
The usage of quantile-based coding has appeared in the literature~\citep{huDeeprETAETAPostprocessing2022,saberianGradientBoostedDecision2019} as a nonlinear feature coding that has demonstrated benefits to model performance in certain problems. 
Generally, we retained only the quantized features in specifying the models except when otherwise specified.
The total size of the feature vector after dropping all original non-quantized numerical features and all constant features expanded to $p=3143.$

\noindent \textbf{Survival Modeling: }
For flexibly modeling the wait time distribution $f$, we use the piecewise exponential survival regression model (PEM)~\citep{friedmanPiecewiseExponentialModels1982}. 
PEMs are defined by specifying the time-dependent hazard $\lambda(t):\mathbb{R}^+ \to \mathbb{R}^+$ using a piecewise constant function, where the hazard changes across breakpoints that define disjoint time intervals.
The probability density function for the PEM follows 
\begin{equation}
    f(t) = \lambda(t) e^{-\int_0^t\lambda(u)\dd u}.
\end{equation}
In this manuscript we set the breakpoints between time intervals at 1 week, 4 weeks, and 9 weeks after discharge.
For each episode $n$, we can estimate a wait time distribution by estimating the log-hazard within each time interval $i,$
\begin{equation}
    \log \lambda_{ni} = \alpha_{ni} + \boldsymbol\beta_{ni}' \boldsymbol{x}_n + \boldsymbol\gamma_{ni}' \boldsymbol{I}_n,
    \label{eq:pem}
\end{equation}
where we allow the model parameters to vary across the data regionally, in a manner that emulates the type of nonlinearity seen in ReLU-nets.
In Eq.~\ref{eq:pem} we separate out the discharge placement effects ($\boldsymbol\gamma_n$) from other effects ($\boldsymbol\beta_n$).
Doing so makes it easier to structure the model for causally interpreting the discharge assignment effects.
We incorporate domain knowledge by acuity-ordering the interventions, enforcing monotonicity of intervention effect by constraining the last five coefficients of $\boldsymbol\xi_n$ to non-positivity.

\noindent \textbf{Causal inference:} We model the discharge placement process $g$ using an ordinal logistic regression model, where 
\begin{align}
    I_n | \boldsymbol{p}_n&\sim \mathrm{Categorical}\left(p_{n0}, \ldots, p_{n5}\right) \nonumber\\
    p_{nk} | \boldsymbol{x}_n, \boldsymbol\nu_n, \boldsymbol{\xi}_n &= \Pr(I_n \geq k| \boldsymbol{x}_n, \boldsymbol\nu_n, \boldsymbol{\xi}_n) \nonumber\\
    &\quad- \Pr(I_n \geq k+1| \boldsymbol{x}_n, \boldsymbol\nu_n, \boldsymbol{\xi}_n) \nonumber\\
    \Pr(I_n \geq k| \boldsymbol{x}_n, \boldsymbol\nu_n, \boldsymbol{\xi}_n) & = \textrm{logit}^{-1}\left(\nu_{nk} + \boldsymbol{\xi}_n' \boldsymbol{x}_n \right),
\end{align} \normalsize
where $\boldsymbol\xi_n$ are slopes corresponding to episode $n$  and $\boldsymbol\nu_n = [\nu_{n1},\ldots,\nu_{n5}]$ are intercepts  
under the constraints $\nu_{nk} < \nu_{n, k+1}$, $\forall k, n$.
The predictions given by this model then feed back into the prediction of the wait time through a slope term for each element of the covariate vector $\boldsymbol{I}_n = \big[ \Pr(I_n\geq 1|\ldots), \ldots,\ \Pr(I_n\geq 5|\ldots),\ 1_{{I_n}\geq 1},\ \ldots, 1_{{I_n}\geq 5} \big]$.
Utilizing the discharge placement probabilities as  model covariates adjusts for the confounding bias caused by the selection process, in a manner analogous to incorporating the local treatment probability as a covariate~\citep{bafumiFittingMultilevelModels2007}.
Additionally, directly modeling the treatment effects within a multilevel model allows us to infer locally-varying treatment effects, partially pooled for stable inference in  regions where the data is sparse ~\citep{gelmanMultilevelHierarchicalModeling2006,fellerHierarchicalModelsCausal2015}.

\noindent\textbf{Parameter decomposition:} The piecewise linear nature of ReLU-nets, and the observation that neural networks produce learned data representations~\citep{goodfellowDeepLearning2016}, suggest that an approach to mimicking their expressivity within regression models is to allows slopes (and intercepts) to vary across regions of the data. We do so by expressing each of these regionally-varying parameters using an additive decomposition.

\begin{figure}
    \centering
    \includegraphics[width=0.95\linewidth]{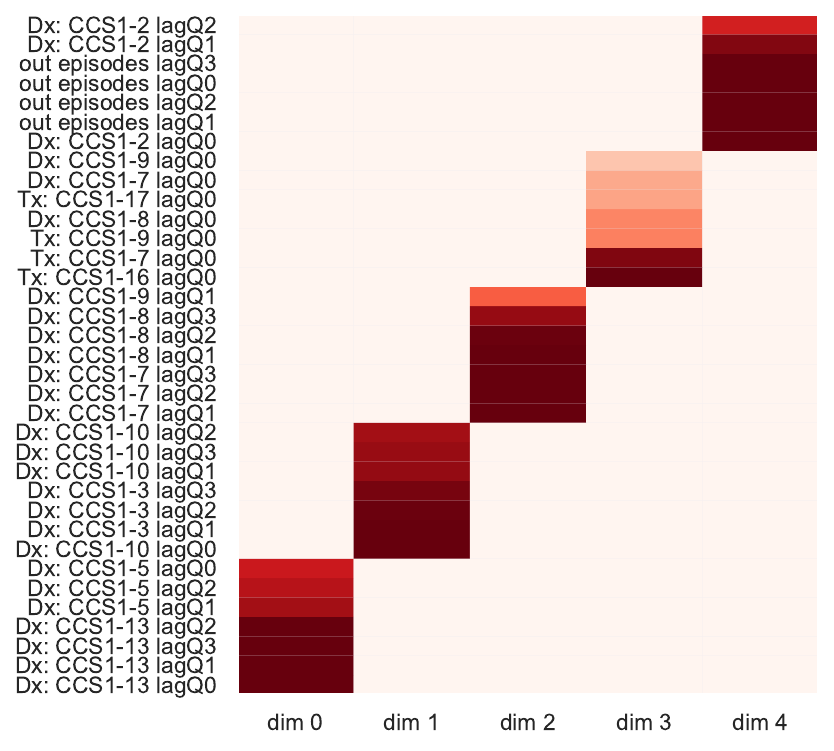}
    \caption{\textbf{History encoding weights for inpatient episodes} where the top seven variables for each dimension of a five dimensional factorization are shown. The features consist of multilevel CCS counts of diagnoses and procedures, as well as counts of the number of episodes, on a quarter-lagged basis. An episodes history representation is found by linear combination of history count features of the given weights.}
    \label{fig:hx_encoding}
\end{figure}

First, for delineating regions in data space (corresponding to cohorts), we project portions of the input data to lower dimensions through unsupervised methods.
In \citet{changSparseEncodingMoreinterpretable2020}, the authors make a connection between sparse probabilistic matrix factorization and probabilistic autoencoders.
We use this approach to develop a low-dimensional representation of the portions of the input covariate vector that pertain to the lagged quarterly history.
Then, we compute the statistics of the learned representation in the training data and develop for each dimension a set of cut-offs to use for bucketization.
This procedure puts each inpatient episode into a specific cohort, represented by a location within a multidimensional lattice, based on medical history.
Specifically, we used a single cut-off (the median) for each of five dimensions (Fig.~\ref{fig:hx_encoding}), creating a set of $2^5=32$ groups based on history.
By design, the rules governing the group assignment can be easily converted to a set of inequalities over sparse subsets of the original data features.
Additionally, we included interactions between the history groups with other discrete attributes such as the major diagnostic category (MDC),  complication or comorbidity (CC) or a major complication or comorbidity (MCC), and race, to create high dimensional discrete lattices where the cells define coarse interaction cohorts in the data.
When partitioning data by a high-order interaction, a big data problem quickly becomes many small data problems -- divide-and-conquer approaches can suffer from overfitting. 
To combat this issue, we developed a multiscale modeling approach where higher-order interactions are regularized by partially pooling their effects into related lower-order interactions.
Specifically, given a multidimensional lattice that represents all cohorts for which the parameter will vary, we assign for each parameter a value within the lattice by decomposing the value into the form 
\begin{align}
    \theta^{(\boldsymbol\kappa)} &= \overbrace{\theta^{(\ast,\ast,\ldots,\ast)}}^{\textrm{\scriptsize zero order}} + \overbrace{\theta^{(\kappa_1,\ast,\ldots,\ast)} + \theta^{(\ast,\kappa_2,\ast,\ldots,\ast)} + \ldots}^{\textrm{\scriptsize first order}} \nonumber\\
    & + \overbrace{\theta^{(\kappa_1,\kappa_2,\ldots,\ast)} + \theta^{(\kappa_1,\ast,\kappa_3,\ast,\ldots,\ast)} + \ldots}^{\textrm{\scriptsize second order}} + \textrm{H.O.T.},
\end{align}\normalsize
where $\boldsymbol\kappa = (\kappa_1, \kappa_2, \ldots, \kappa_D)$ is a $D$ dimensional multi-index.
In practice, we truncate the maximum order of terms in this decomposition due to memory constraints.
More details on the exact decompositions that we used for our model parameters can be found in the Supplemental Materials.

\noindent\textbf{Regularization:} By design, the parameter decomposition method inherently regularizes by partial pooling~\citep{gelmanMultilevelHierarchicalModeling2006}.
Additionally, we used weakly informative priors on the component tensors in these decompositions in order to encourage shrinkage at higher orders.
For the regression coefficients, we utilized the horseshoe prior for local-global shrinkage
\citep{ghoshModelSelectionBayesian2017,bhadraHorseshoeRegularizationMachine2019,polsonShrinkGloballyAct2011,vanerpShrinkagePriorsBayesian2019}.
Please see the Supplemental Materials for more details on the model specification.

\noindent\textbf{Implementation:}
We used Tensorflow Probability~\citep{dillonTensorFlowDistributions2017}, developing a set of libraries for managing the parameter decompositions that is publicly available at \texttt{github:mederrata/bayesianquilts}. 

We trained our model using minibatch mean-field stochastic ADVI, using batch sizes of $10^4$, and a parameter sample size of $8$ for approximating the variational loss function.
We utilized the Adam optimizer with a starting learning rate of $0.0015$, embedded within a lookahead optimizer~\citep{zhangLookaheadOptimizerSteps2019} for stability.
Each epoch where the mean batch loss did not decrease, we set the learning rate to decay by 10\%.
Training was set to conclude if there was no improvement for 5 epochs, or if we reached 100 epochs, whichever came sooner.
More information on the training is present in the Supplemental Materials.
We used scikit-learn 1.1.1 for fitting baseline logistic regression models, and 
XGBoost 1.6.1 for fitting a reference blackbox model for comparison.
We implemented a horseshoe Bayesian convolution neural network with ReLU activation using TFP, where we used a single hidden layer of size one-fifth the input layer.
For computing global SHAP values, we used regression-based KernelSHAP~\citep{covertImprovingKernelSHAPPractical2021}.
All computation was performed using the Pittsburgh Supercomputing Center's Bridges2 resources.
We utilized extreme memory (EM) nodes for preprocessing, and Bridges2-GPU-AI for training.

\section{Results}

\begin{table}
\centering
\begin{tabular}{lcll} 
\toprule
Model & Interpretability  & AUROC & AUPRC  \\ \hline
XGBoost   & None                     & 0.758 & 0.497  \\
ReLU-BNN   & Computationally                    &   0.757    &   0.500     \\
LR w/o quantization  & Comprehensibly & 0.731 & 0.381 \\
LR    & Comprehensibly                    &   0.755    &    0.481   \\
\textbf{PEM} & \textbf{Mechanistically} & \textbf{0.757}  & \textbf{0.497} \\
\bottomrule
\end{tabular}
\caption{
\textbf{30-day unplanned readmission or death classification metrics} for evaluated models: XGBoost, Sparse logistic regression (LR), Bayesian neural network (BNN), our Piecewise exponential model (PEM).  Quantization refers to the histogram-based bucketization of real-valued features. Area under the receiver operator curve (AUROC) and area under the precision-recall curve (AUPRC) computed on held-out 2012 inpatient episodes. Models trained on 2009-2011 episodes. Interpretability judged according to Fig.~\ref{fig:interpretability}.
}\label{tab:auc}
\end{table}

\textbf{Prediction Accuracy:} Table.~\ref{tab:auc} shows the classification accuracy of our model in predicting readmissions or death within the first 30 days, benchmarked against predictions given by alternative models trained on the same dataset.
The standard deviation in both the AUROC and AUPRC measures, as determined using bootstrap, was approximately $0.003$.
Non-linearly transforming our count features using quantization improved the accuracy of logistic regression to nearly match that of XGBoost on this dataset as measured by AUROC.
Hence, we used quantization for features in both the Bayesian neural network (BNN) and piecewise exponential  (PEM) models.
The Bayesian neural network we developed utilizes sparsity-inducing horseshoe priors~\citep{carvalhoHorseshoeEstimatorSparse2010} on the weights and biases, which has been shown to improve model performance~\citep{bhadraHorseshoeRegularizationMachine2019}.

\noindent\textbf{Interpretation:}
In addition to being competitive with blackbox methods in terms of prediction accuracy, our model, as a generalized linear survival regression model, is easily interpretable.
To be specific, our model is a generalized linear survival model where the coefficients vary. The value of each coefficient is the logarithm of a hazard ratio corresponding to the effect of a given feature, for a given data cohort, for a given time period. Log hazards greater than zero correspond to increased probability of event (readmission or death).
Here, we provide select portions of the \emph{ground-truth} global interpretation of the model, found by simply reading off the values of the regression coefficients. Please see the Supplemental Materials for a more-complete accounting of the model. This type of exposition is impossible with blackbox models without relying on unreliable approximations. 

\begin{figure}[]
    \centering
    \includegraphics[width=0.9\linewidth]{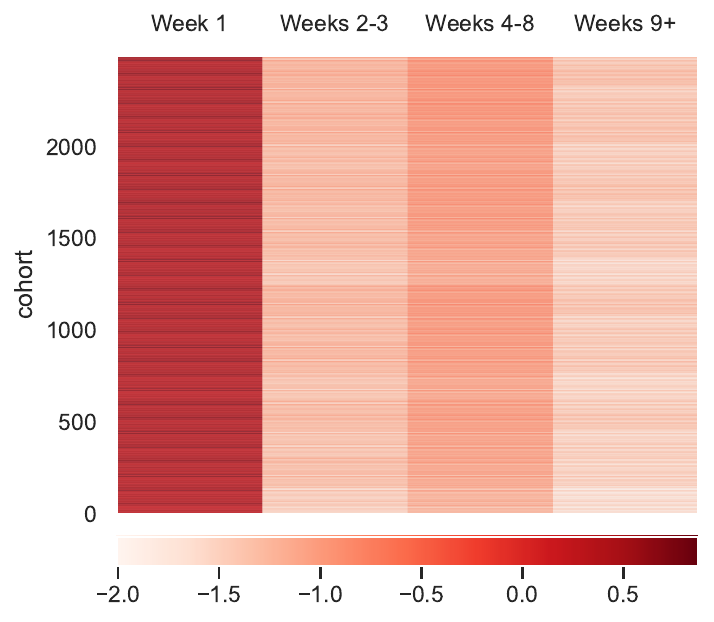}
    \caption{\textbf{Mean baseline log-hazards by week for each episode interaction cohort defined within the model.} Larger log-hazards corresponds to more readmission risk. Personalized values of $\boldsymbol{\alpha}_n$ specific to each episode are found by mapping an episode into its cohort grouping.  }
    \label{fig:baseline_survival}
\end{figure}
\noindent\textbf{Time-dependent risk factors:} The model segments the data based on low-dimensional representation and assigns for each predictor a cohort-level effect within each time interval.
The cohorts are delineated by the recent history of medical services utilization and the properties of the present hospital admission.
The effects within the model are hazard ratios, which describe the instantaneous relative risk associated with a predictor relative to a baseline.
In most cases, the baseline refers to a typical or normal value of a variable.
Membership to cohorts also itself is associated with a baseline risk -- baseline log-hazards are presented in Fig.~\ref{fig:baseline_survival} for the $12480$ episode cohort types defined within the  decomposition for the parameter vector $\boldsymbol{\alpha}_n$ in Eq.~\ref{eq:pem}.
Larger values of the hazard imply higher probability of event (readmission or death).
There exists variability in the hazards across cohorts (rows), though the most striking change is in time (columns).
Generally, the hazard is greatest in the first week after discharge.
This finding implies that patients are more vulnerable in the first week than afterwards -- keeping a patient out of the hospital within the first week has the largest impact on the overall risk that they will die or be readmitted.
For this reason, we will focus on understanding the model's predictions of the first-week risk.
\begin{figure}[]
    \centering
    \includegraphics[width=\linewidth]{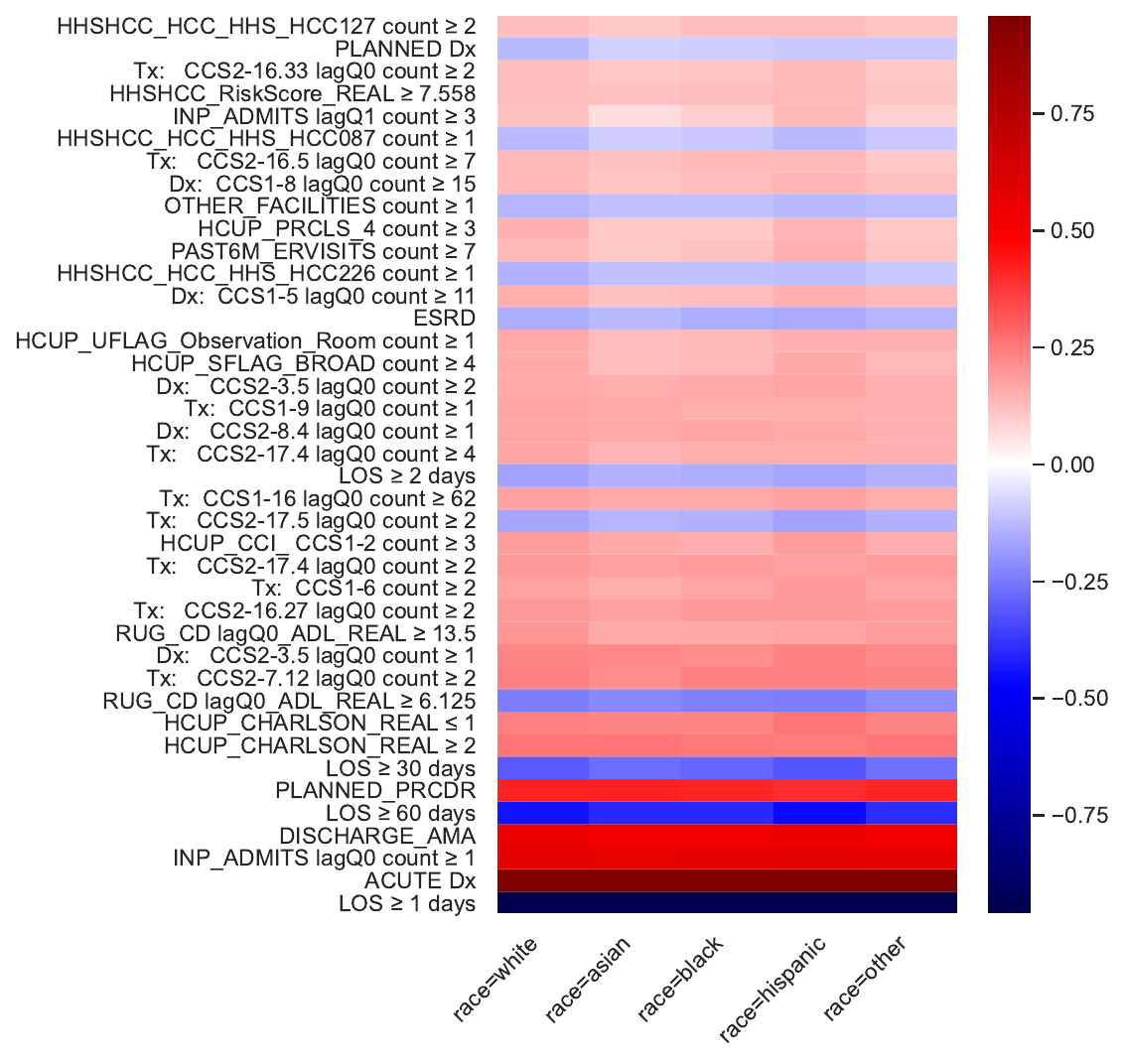}
    \caption{\textbf{The 40 predictors with the largest absolute coefficients in the first week (through day 7) after readmission.} All predictors are binary and all parameters are additive log hazard ratios. Higher (red) corresponds to larger hazards and greater readmission risk.}
    \label{fig:week1}
\end{figure}

The $40$ most-impactful first-week factors are shown in Fig.~\ref{fig:week1},
where the parameters have been decomposed in order to control for racial biases.
The most-predictive single feature was length of stay.
Lengths of stay less than a full day
had a relative log hazard ratio of 0.97 (95\% CI: 0.96 -- 0.98) (note LOS$<$1 day was the reference group and so is the converse of LOS $\geq$ 1 days shown in Fig.~\ref{fig:week1}).
Having an acute primary diagnosis code, at least one inpatient stay in the previous quarter (lagQ0, within 90 days of admit), and discharge against medical advice were also strong predictors associated with increased risk of readmission or death.
Patients who received skilled nursing care in the quarter preceding an inpatient episode, who had a Resource Utilization Group (RUG) Activities of Daily Living (ADL) score of at least $6.125$ tended to have a lower risk of readmission in the first week than otherwise, however, the risk increased for quarter-lagged ADL scores of at least $13.5.$

\begin{figure}[]
    \centering
    \includegraphics[width=0.95\linewidth]{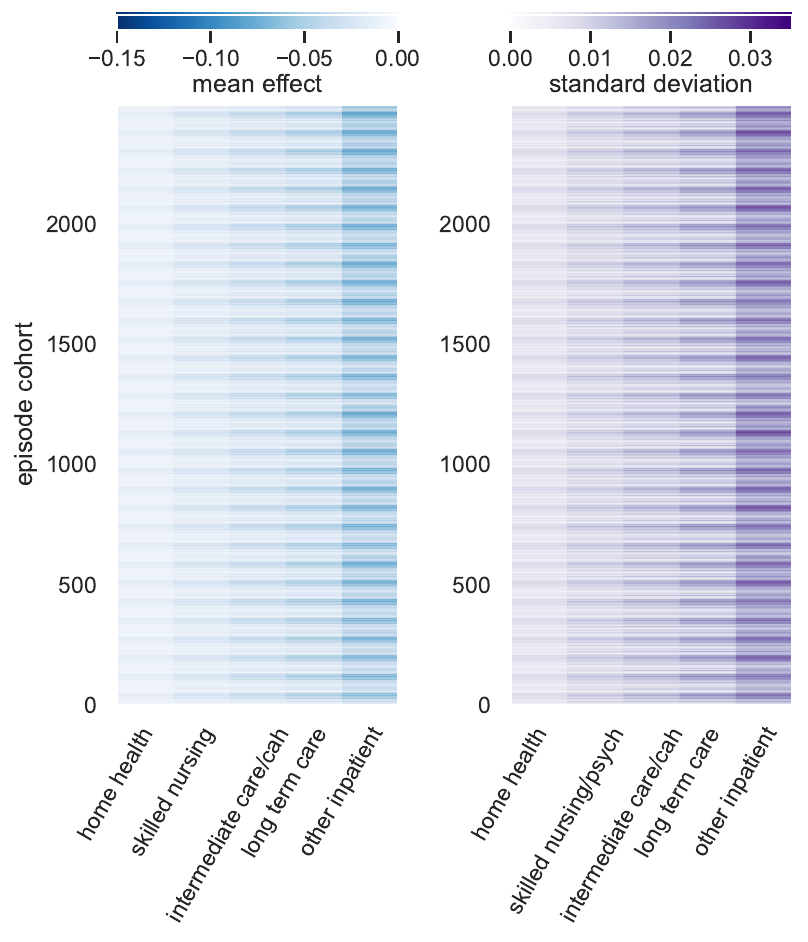}
    \caption{\textbf{First-week effects of discharge placement:} Mean (left) and standard deviation (right) by cohort (row) of the five placement interventions assessed, in increasing order of implied acuity. Effect is difference in log-hazard relative to a normal discharge (home).}
    \label{fig:effects}
\end{figure}

\noindent\textbf{Discharge placement effects:} In Fig.~\ref{fig:effects}, we show the cohort-wise causally-adjusted mean local average treatment effects of discharge to each of the given care settings as well as the local standard deviation in the effect. Focusing on the effect of discharging to skilled nursing care, the effects were greatest for episodes graded by DRG code as having either a complication or comorbidity (CC) or a major complication or comorbidity (MCC). In particular, CC/MCC episodes with a major diagnostic code of 2 (Diseases and Disorders of the Eye), 14 (Pregnancy, Childbirth And Puerperium), and 22 (Burns) have the greatest response to discharge to skilled nursing.

\begin{figure}[]
    \centering
    \includegraphics[width=0.95\linewidth]{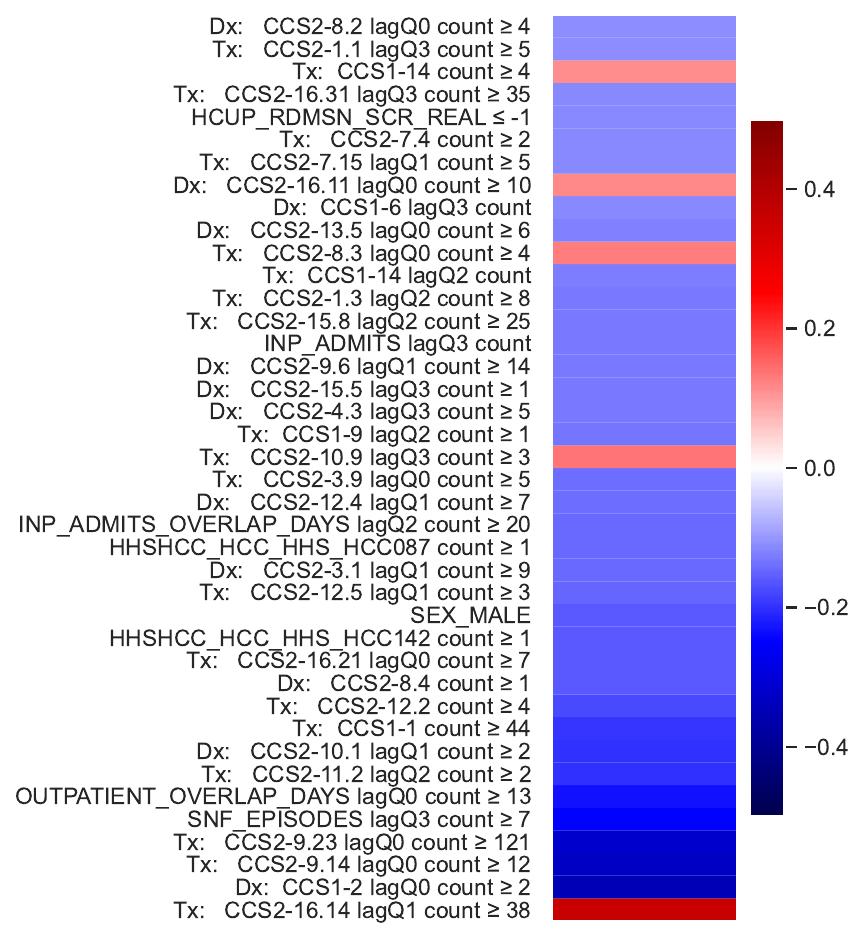}
    \caption{ \textbf{Shapley values:} Top 40 absolute feature weights for prediction of $30$ day readmission using our survival model, where we have the ground truth explanation (See Figs~\ref{fig:hx_encoding}--\ref{fig:effects}, and the supplement for how the features actually are incorporated into the model. SHAP fails to identify the features a model is using whenever features are correlated.}
    \label{fig:shap}
\end{figure}

\noindent \textbf{Posthoc-xAI (SHAP) misleads:} Knowing what the model is doing in exact terms, let us see how posthoc-xAI thinks the model is working.
In Fig.~\ref{fig:shap} we display the most important model features as determined by magnitude of global SHAP values in the prediction of readmission or death within the first 30 days.
 SHAP is  computationally costly to approximate -- the details of our SHAP computation are available in the Supplemental Materials.
The four most-influential features according to the explainer are specific CCS classes of treatments and diagnoses in the recent quarterly history.
Comparing these results to the parameter values of Fig.~\ref{fig:week1}, it is evident that the feature sets disagree.
Nor do the values in Fig.~\ref{fig:shap} align with parameter values for later weeks (see Supplemental Materials).
This finding is unsurprising; SHAP has been consistently shown to fail to recover ground-truth interpretations~\citep{kumarProblemsShapleyvaluebasedExplanations2020,bordtPostHocExplanationsFail2022}, in problems where the predictors are correlated. 
SHAP fundamentally does not answer the question of what a given model is doing in order to reach a prediction.
Furthermore, feature importance is not grounded in any relevant units and also does not speak to relevant interactions that are captured in a model.
We criticize SHAP because it is one of the most popular posthoc-xAI techniques, however, similar arguments hold for other techniques~\citep{rudinStopExplainingBlack2019,babicBewareExplanationsAI2021,zhouFeatureAttributionMethods2022}.

\section{Discussion}

We presented a method for mimicking ReLU-nets within inherently interpretable multilevel Bayesian models.
We applied this methodology to the prediction of hospital readmissions or death after discharge, and to the causal inference of the effects of discharge assignments.

\noindent\textbf{Accuracy without blackboxes: } We demonstrated how we were able to perform like blackboxes, without sacrificing interpretability.
We accomplished this feat through two classes of methods: First, our novel modeling framework allowed us some fine-grain resolution in looking at the differential effects of the predictors in data subgroups. Additionally, it helped regularize the inference of local average treatment effects for choosing discharge placement. Second, we performed layers of feature engineering. The first layer was an extraction of medically-relevant information from the raw billing that gave us attributes such as chronic diseases, comorbidities, and ADL function.
Then, we reduced noise in the raw coding by mapping to the clinically-relevant CCS system. These two steps were sufficient for our logistic regression model to match the performance of an XGBoost model in the literature based on the same dataset~\citep{mackayApplicationMachineLearning2021}. Finally, we performed feature quantization based on the per-feature statistics. Quantization led to a big performance increase in logistic regression and also in the neural network for a given model size. We took these lessons and used them in defining our interpretable survival model.

\noindent\textbf{Posthoc xAI is inherently untrustworthy: } Our model, being a regression model is inherently interpretable. It admits an unequivocal ground-truth explanation that is found by simply examining its regression coefficients, all of which are log hazard ratios. Hence, it is a good test case for testing the accuracy of posthoc explainers. We tested SHAP on our model; it failed in coming close to the ground-truth.
This finding is consistent with other literature that has looked critically at SHAP and other xAI tools.

While posthoc-xAI does not make blackboxes interpretable,  interpretability is not always unnecessary. Quantifying sample average treatment effects and making predictions does not require interpretable modeling~\citep{hillBayesianNonparametricModeling2011}, or even necessarily models at all~\citep{dingModelfreeCausalInference2017}.
Blackbox methods offer good performance with minimal thoughtfulness.
For these reasons, blackbox methods remain inherently useful -- so long as one does not whitewash them with false explainability.

\noindent\textbf{Limitations:} Our modeling approach has downsides. Numerical stability generally requires the use of double precision floating point. The lattice-based parameter decomposition is memory-intensive which in some applications may severely limit expressivity.

\section{Acknowledgments}
We thank the Innovation Center of the Center for Medicare and Medicaid services for providing access to the CMS Limited Dataset through DUA LDSS-2019-54177. We also thank Dr. Pei-Shu Ho for help in understanding Medicare billing data. CCC is supported by the Intramural Research Program of the NIH, NIDDK. This work used the Extreme Science and Engineering Discovery Environment (XSEDE)~\citep{xsede}, which is supported by National Science Foundation grant number ACI-1548562  through allocation TG-DMS190042. 


\Urlmuskip=0mu plus 1mu\relax
\bibliographystyle{plainnat}
\bibliography{mederrata}


\newpage

\clearpage

\appendix

\renewcommand\thefigure{S\arabic{figure}}    
\setcounter{figure}{0}   

\section{Supplementary Methods}

\subsection{Medicare data preprocessing}

Here we describe some details on the choices we made in preprocessing that will help make our work reproducible. Kyle Barron's \href{https://kylebarron.dev/medicare-documentation/}{Medicare Documentation} repository of Medicare data documenation is an excellent resource for acquainting oneself with this standardized dataset.
Our first steps in processing the CMS LDS were to merge the files, originally organized by year, into long tables for each claim type.
In the process, we renamed pre-2011 columns in the dataset to match 2011+ plus columns where-ever they differed. We will refer to the dataset using 2011 and beyond column names.

\subsubsection{Episode Grouping}

The CMS LDS consists of records organized into claims. Multiple claims can constitute a single period or episode of service. We determined episodes of the following types:
\begin{enumerate}
    \item inpatient (inp)
    \item skilled nursing facility (snf)
    \item hospice (hosp)
    \item outpatient (out, car)
\end{enumerate}
For determining episodes, we grouped claims of each of the given types by person, and sorted by either the admission date (for inp, snf, hosp), or the claim through-date for (out, car). 

Then for inp, snf, hosp, we merged successive claims into running episodes if they overlapped temporally, if the provider was the same and the intermediate discharge code indicates that the individual was not otherwise discharged home in between (we allow for distinct episodes with zero days of wait if a patient is discharged home and returns on the same day).

For out and car, we did the same merging with all claim types together, relaxing the need for the provider to match in an episode. Then we filtered for out/car episodes that did not overlap with inp, snf, hosp episodes -- we determined these to be true outpatient episodes.

Then, for out and inp episodes, we determined if they corresponded to emergency department visits by looking for corresponding revenue center codes.

\subsection{Model Specification}

\begin{table}[h]
    \centering
    \begin{tabular}{c c  c}
    \toprule
        Parameter &  Decomposition
         &  Max order \\ 
         \hline
         $\boldsymbol\alpha$ & MDC  $\times$ Hx $\times$ CC/MCC & 2 \\
         $\boldsymbol\beta$ & race & 1 \\
         $\boldsymbol\gamma$ & MDC $\times$ Hx $\times$ CC/MCC & 2\\
         $\boldsymbol\nu$ & MDC $\times$ Hx $\times$ CC/MCC & 2\\
         \bottomrule
    \end{tabular}
    \caption{\textbf{Specific decompositions used per parameter to define cohorts}, where major diagnostic category (MDC) is of size 26, history (Hx) is of size $2^5$, corresponding to low/high in each of the five dimensions, CC/MCC is of size $3$, and race is of size $5.$}
    \label{tab:decompositions}
\end{table}

The specific decompositions that we used for each of the model terms are displayed in Fig.~\ref{tab:decompositions}. For the missing parameter $\boldsymbol\xi,$ the results in this manuscript are all determined using $\boldsymbol\xi=\boldsymbol0.$

The python package \texttt{bayesianquilts}, with demonstration available at \texttt{github:mederrata/bayesianquilts} contains utilities for managing decompositions such as these.

We used a regularized horseshoe prior in order to encourage $\boldsymbol\beta$ to be sparse.
Specifically, we applied an independent horseshoe prior to this parameter within every model cohort.

The individual components of each of the parameter decompositions were all modeled using Gaussian weakly-informative priors (with a default scale of 5 for the zero-order terms in the expansion). 
We helped encourage shrinkage by having the scale of these priors decay for higher order terms in the decomposition. For the results in the paper, we used a decay factor of $0.1$ per each order.

\subsubsection{Training}

We utilize TFP's ADVI routines, which utilize stochastic sampling in computation of the ELBO. 
For this reason, it is not uncommon for specific parameter combinations to be in highly improbable locations -- which can trigger underflows.
To avoid instabilities, re adjust the likelihood on a per-observation level, first computing the minimum finite value of the log likelihood and then setting any divergent values to the minimum finite value minus a fixed offset of $100.$
We use the soft-plus function as a default bijector for any parameters that are supposed to be non-negative. 

\subsubsection{SHAP}

KernelSHAP, the general all-purpose model-agnostic implementation of SHAP, is very resource intensive, so we had to tune it in order to run it.
First, we used the regression-based version of KernelSHAP, found in the python package \texttt{shapreg}~\citep{covertImprovingKernelSHAPPractical2021}, which is not as resource intensive.

Second, although we had a very powerful computational resource at our disposal (Pittsburgh Supercomputer Center Bridges 2-AI with 512GB RAM), we had to restrict the input data size to 5k random training examples.
Otherwise, we found that the system would run out of memory, causing the application to segmentation fault.

Finally, in order to get around an error involving a singular matrix, we regularized the linear algebra problem embedded within the algorithm, adding a fixed small constant of $10^{-8}$ to the diagonal of the linear transformation matrix (see \href{https://github.com/iancovert/shapley-regression/compare/master...joshchang:shapley-regression:mederrata}) for the exact modification made.

We were able to run \texttt{shapref} with \texttt{n\_samples=2400} and \texttt{batch\_size=24} on our resource in approximately 5 hours. 
Due to the memory requirement issues with our large dataset, we are not able to scale this result to more data.
SHAP is known to be computationally expensive, particularly for large datasets and a large number of features (see github issues \href{https://github.com/slundberg/shap/issues/1053}{1053} \href{https://github.com/slundberg/shap/issues/1495}{1495}), and its very computation is inherently based on approximations.
We believe that our computation of SHAP for our model is a reasonable representation of how well-approximated it is in practice, on a real problem and on a real dataset with a large number of predictors.

Our main point in the main text is a reiteration of the well-known fact that SHAP feature importance is not guaranteed to match what a model is doing in practice when the features in the training data are correlated -- as would be true in most real-world problems. Examining Fig.~\ref{fig:shap} in the context of Fig.~\ref{fig:discharge_effects_all}, one sees that the most-important SHAP features are not themselves very important in the model.

\section{Supplementary Results}

Here are results omitted from the main text for space constraints.

\begin{figure}
    \centering
    \includegraphics[width=\linewidth]{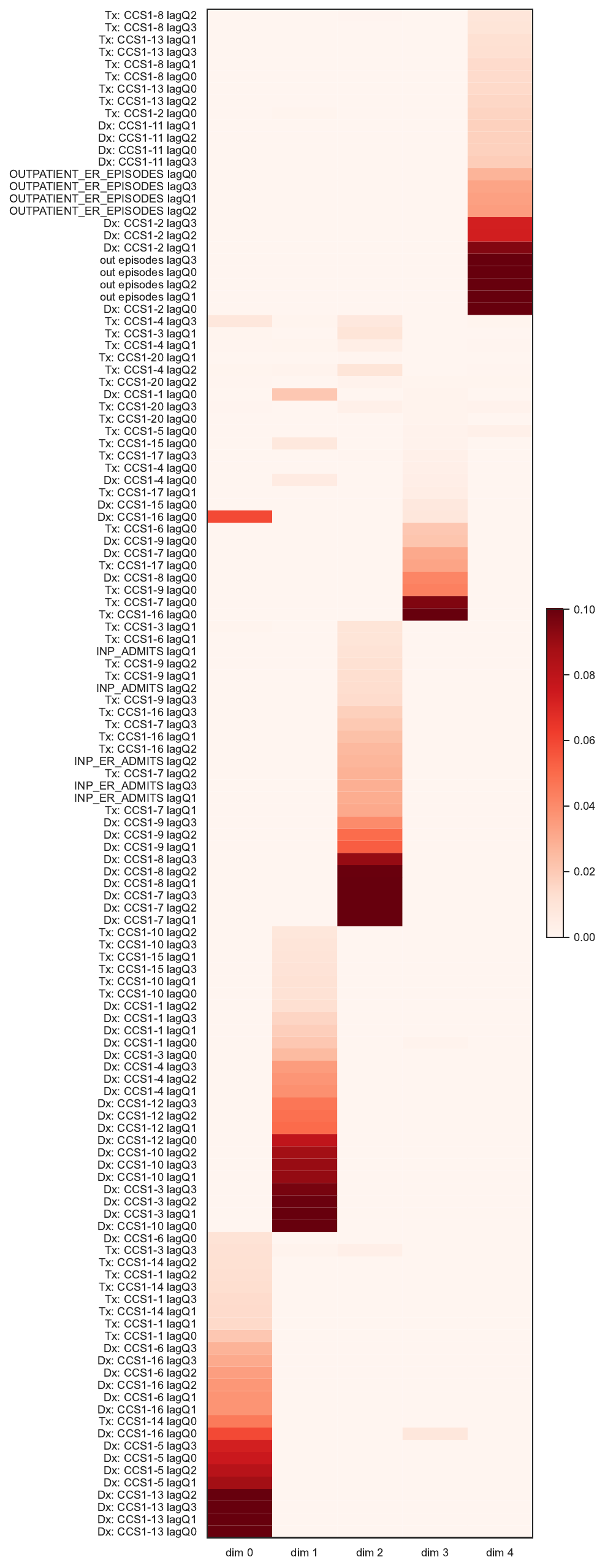}
    \caption{\textbf{Extended version of Fig.~\ref{fig:hx_encoding}} with up to $25$ features per dimension}
    \label{fig:hx_encoding_expanded}
\end{figure}

\begin{figure}
    \centering
    \includegraphics[width=\linewidth]{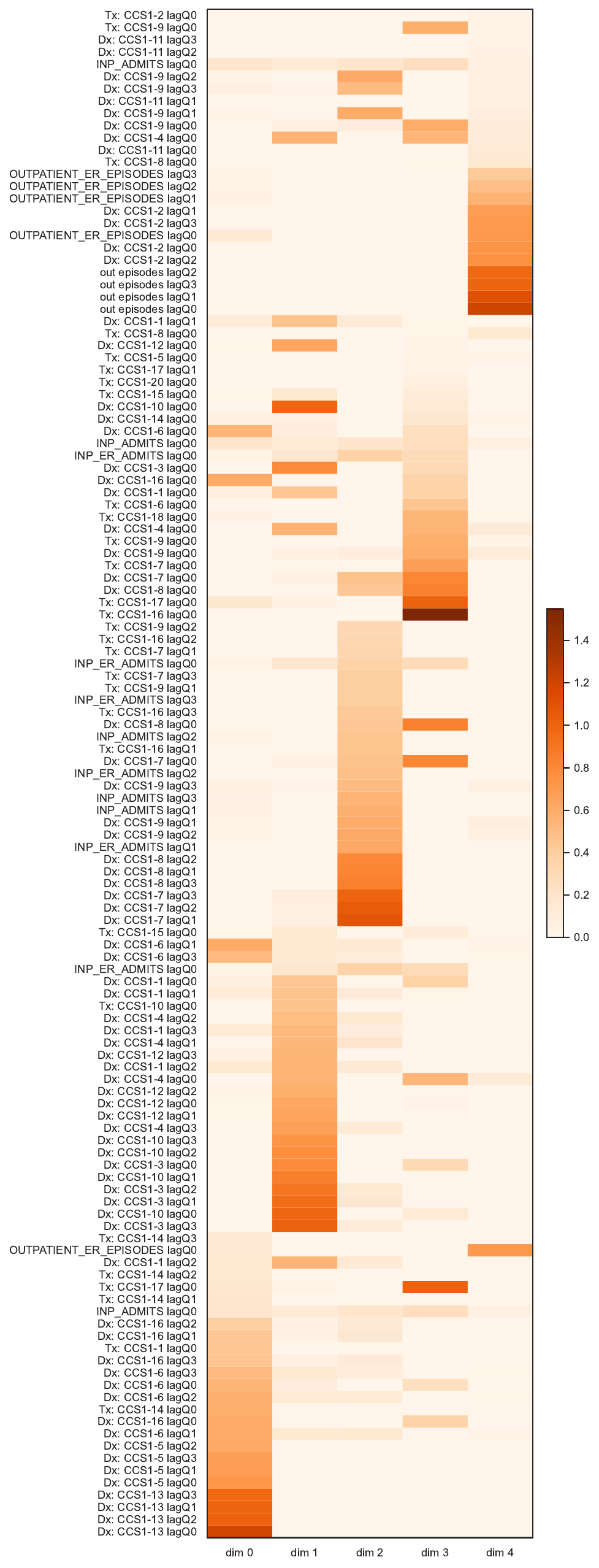}
    \caption{\textbf{Decoding matrix corresponding to the encoding model of  Fig.~\ref{fig:hx_encoding_expanded}} showing up $25$ features per dimension}
    \label{fig:hx_decoding}
\end{figure}

\subsection{History representation}
We utilized sparse probabilistic matrix factorization in order to obtain a low-dimension representation of personal medical history for the year prior to each episode.
The encodings given by the model (Fig.~\ref{fig:hx_encoding_expanded} is an expanded version of Fig.~\ref{fig:hx_encoding} from the main text) specify linear combinations of the original data features that define a representation of an episode's history.
The representations then can be constituted into a predictive distribution for the original features by transformation against a decoding matrix (Fig.~\ref{fig:hx_decoding}).
Note that this method finds a subset of the input features that can be used to predict the value of all features.

\subsubsection{Random slopes}

\begin{figure*}
    \centering
    \includegraphics[width=\linewidth]{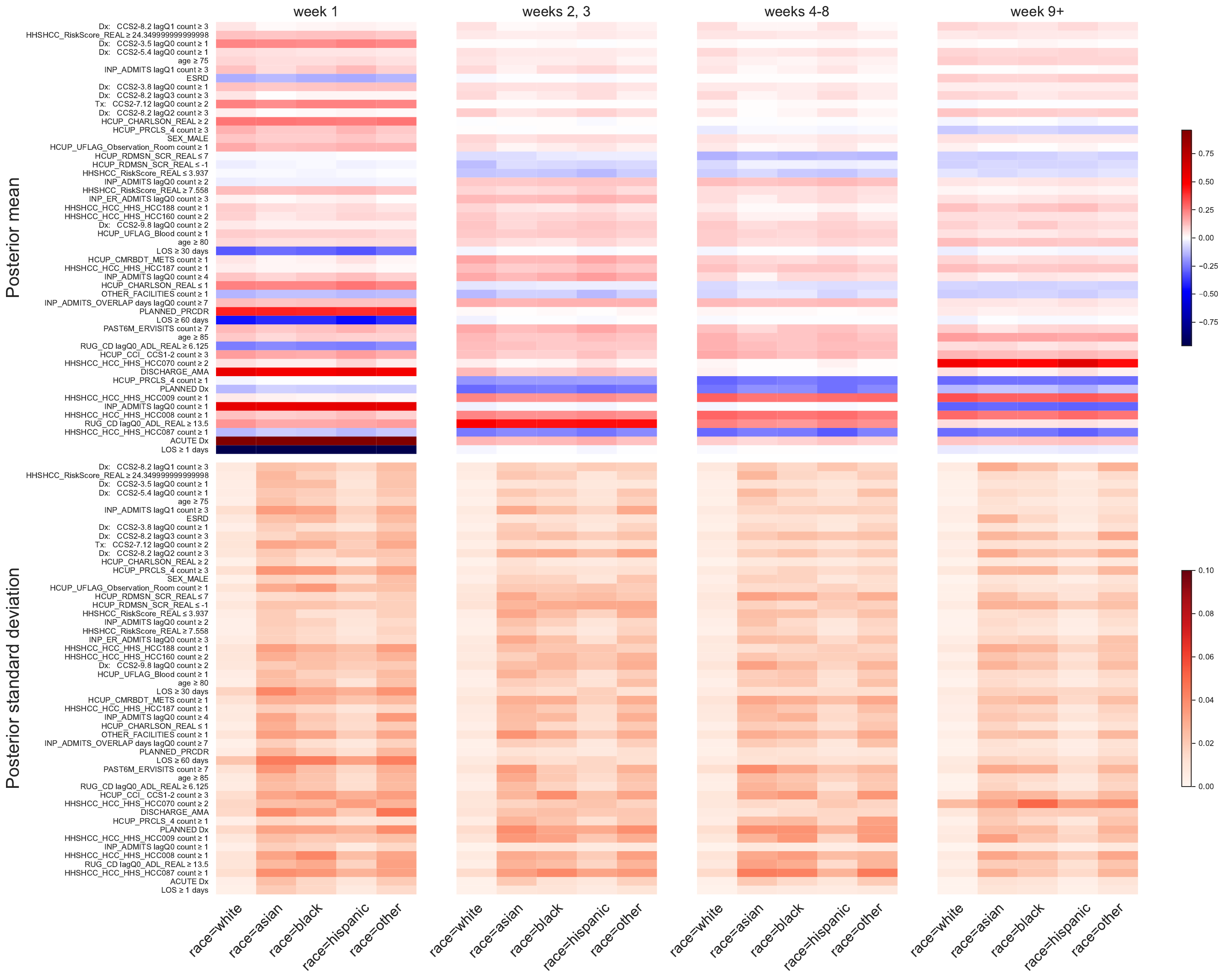}
    \caption{\textbf{The 50 most influential regressors $\boldsymbol\beta$ (posterior mean, standard deviation)} tracked through all time intervals. A more-comprehensive version of this figure can be found in our other supplemental file.}
    \label{fig:beta_all}
\end{figure*}

Although we do not use this terminology in the main text, in the language of hierarchical mixed effects models the parameters $\boldsymbol\beta$, $\boldsymbol\xi$ in the model are random slopes. In the main text we presented the week 1 slopes in Fig.~\ref{fig:week1}. In Fig.~\ref{fig:beta_all}, we present the components of $\boldsymbol\beta$  of the largest magnitudes, across all time intervals.
As we noted in the main text, length of stay being at least 1 day, or conversely, being less than a full day, was the most impact predictor of early readmission.
However, the effect disappears after one week. 
Long length of stay (greater than 30 days) appeared to follow the same trend, with those having a length of stay of at least a month having a lower readmission risk in the first week after discharge, but not reduced risk after the first week.
Generally, the magnitude of the slopes tended to increase over time, with a few exceptions.

 \subsubsection{Random intercepts}
 
 The parameter $\boldsymbol\alpha$ from Eq.~\ref{eq:pem} is specific to each cohort in the model -- it is a random intercept in hierarchical mixed effects modeling terminology. We presented the posterior mean for this parameter in Fig.~\ref{fig:baseline_survival}, interpreting this quantity as a cohort-specific baseline survival. 
 
\subsubsection{Causal inference}

\begin{figure}
    \centering
    \includegraphics[width=\linewidth]{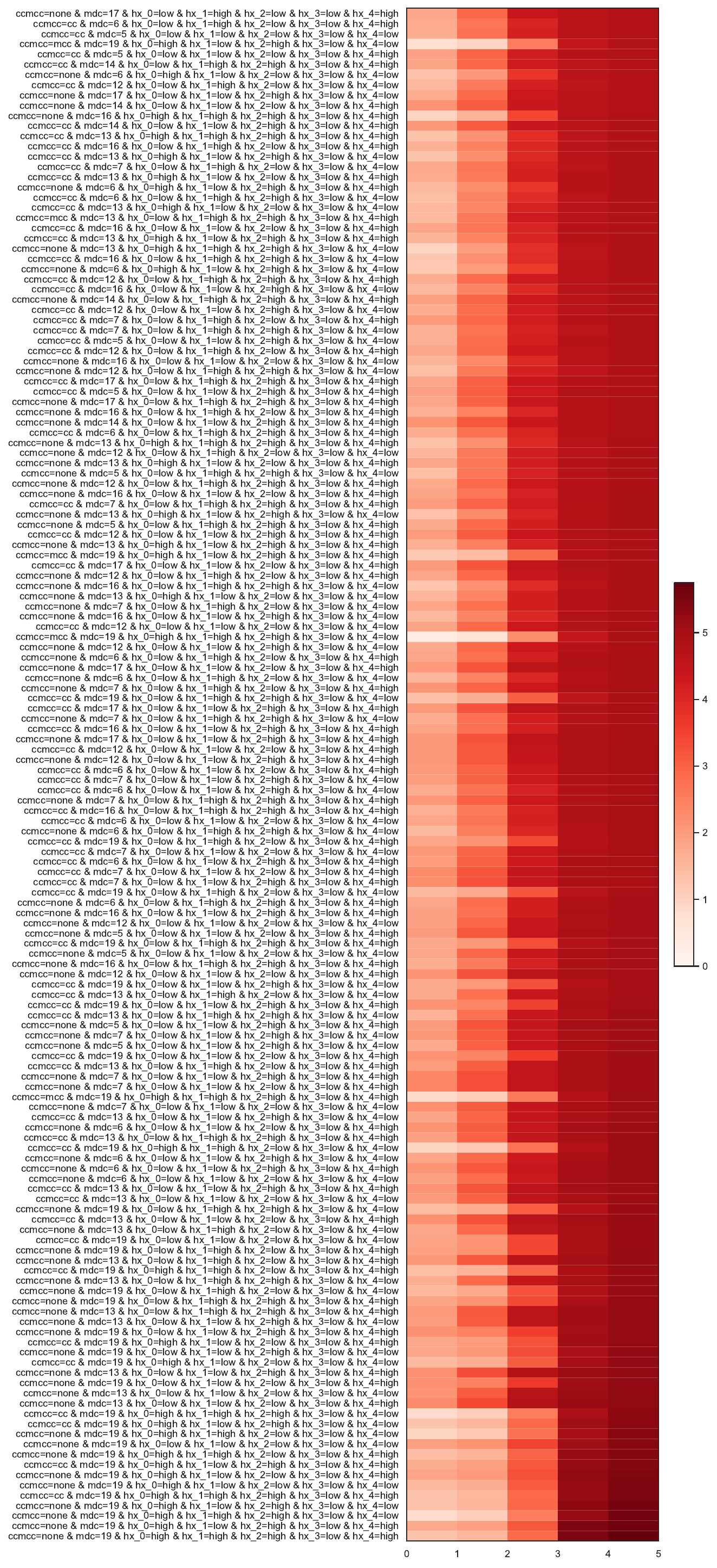}
    \caption{\textbf{Cohort-wise ordinal intercept terms} $\boldsymbol\nu$ for the prediction of the distribution of discharge assignments.}
    \label{fig:nu}
\end{figure}

In our model we adjust for treatment selection bias by incorporating  estimates of the treatment probabilities as covariates. The ordinal logistic regression intercepts are provided in Fig.~\ref{fig:nu}.

\begin{figure*}
    \centering
    \includegraphics[width=\textwidth]{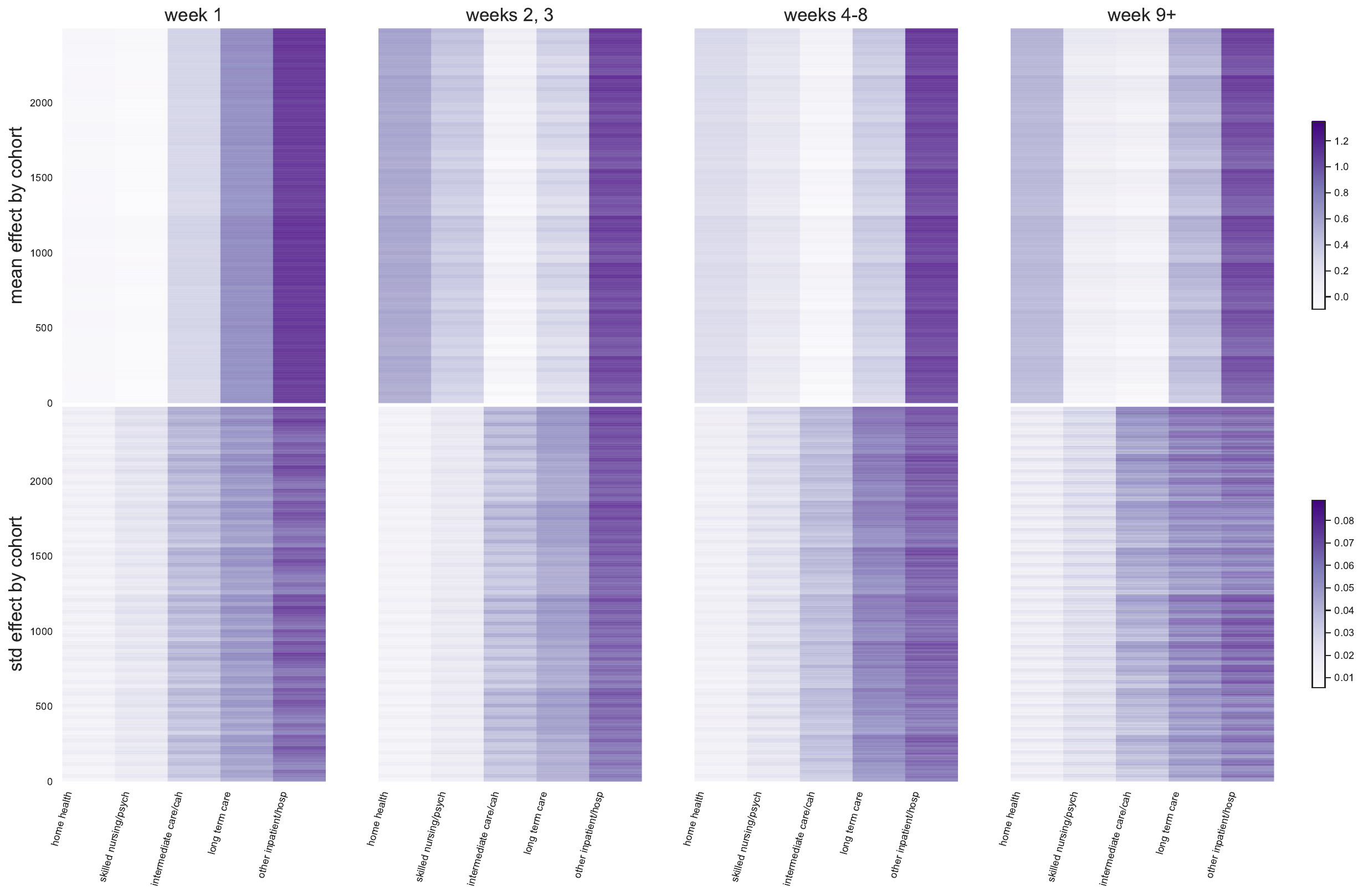}
    \caption{\textbf{Discharge assignment adjustment terms corresponding to first five terms of $\boldsymbol\gamma$ (posterior mean, standard deviation)} in terms of log hazard ratio under the log-additive effects model of Eq.~\ref{eq:pem}}
    \label{fig:adjustment_all}
\end{figure*}

\begin{figure*}
    \centering
    \includegraphics[width=\textwidth]{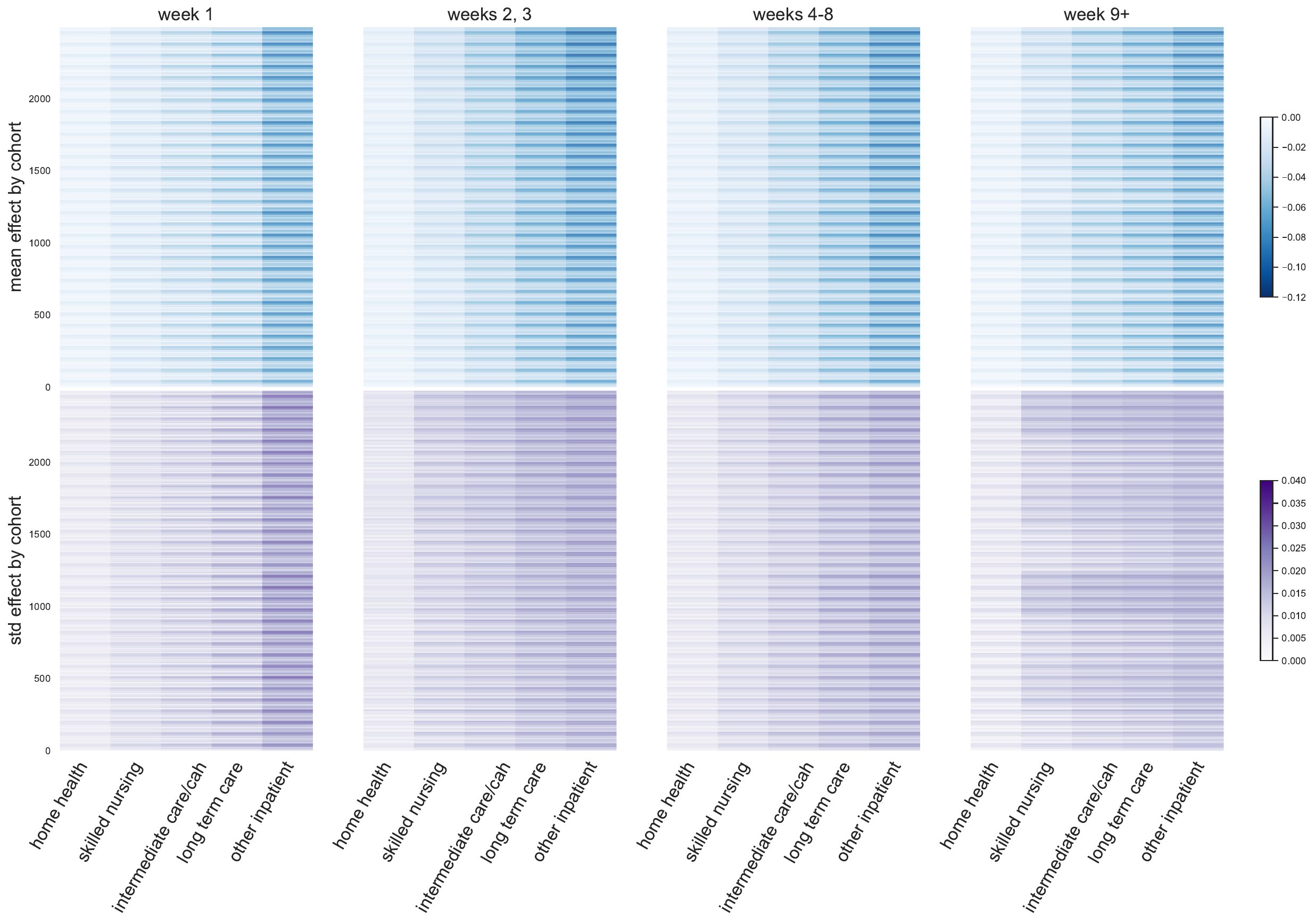}
    \caption{\textbf{Discharge placement effects $\boldsymbol\gamma$ (posterior mean, standard deviation)} in terms of log hazard ratio under the log-additive effects model of Eq.~\ref{eq:pem}}
    \label{fig:discharge_effects_all}
\end{figure*}

\begin{figure*}
    \centering
    \includegraphics[width=\textwidth]{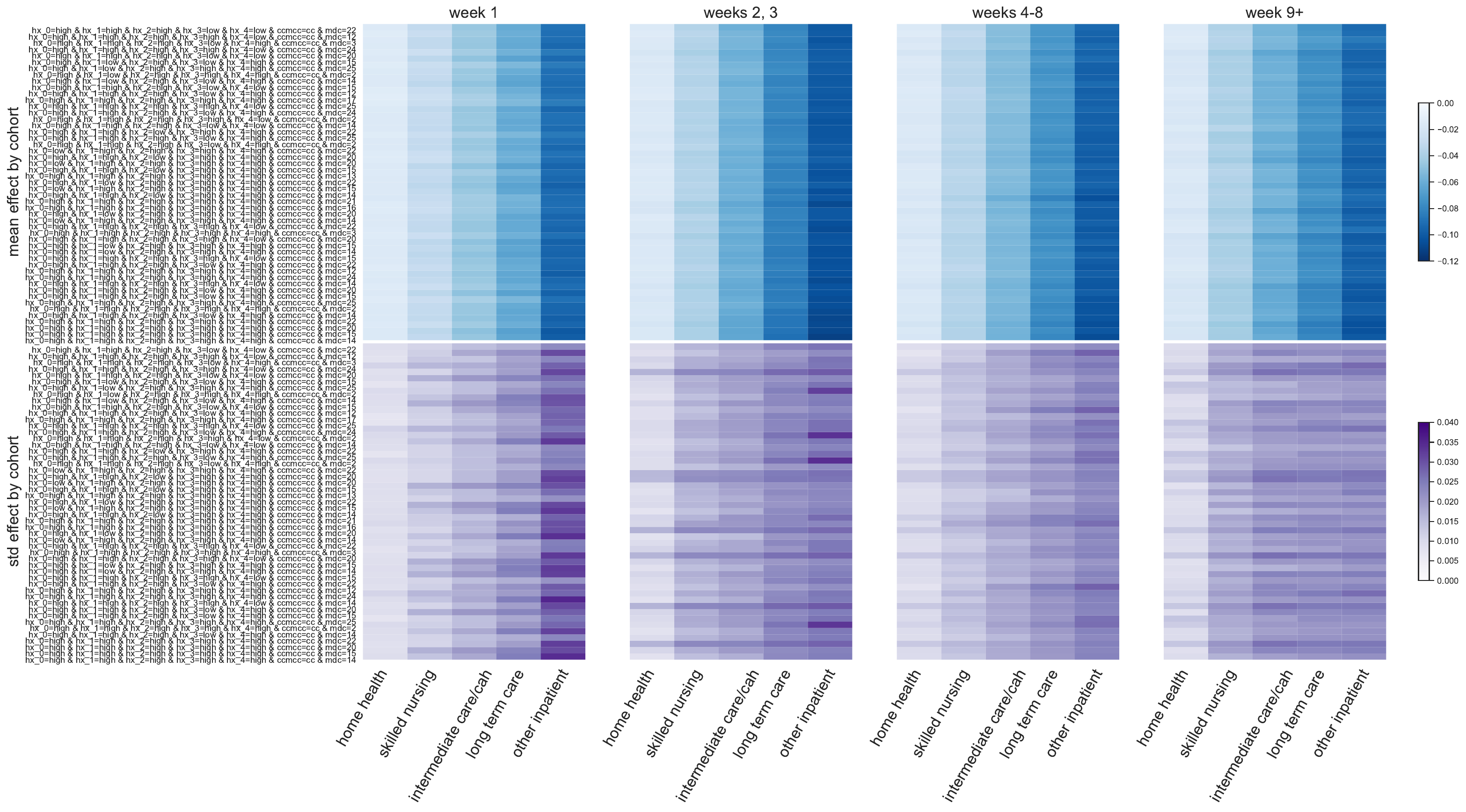}
    \caption{\textbf{Discharge placement effects for select cohorts} with the largest mean discharge placement effects.}
    \label{fig:select}
\end{figure*}

 The first five components in the parameter $\boldsymbol\gamma$ adjust for the selection bias present in claims. We present our cohort-specific estimates of $\boldsymbol\gamma$ in Fig.~\ref{fig:adjustment_all}. We present the full timecourse of discharge placement effects in Fig.~\ref{fig:discharge_effects_all}.
 Largely, the discharge placement affects appear to strengthen from week 1 to weeks 2/3 before weakening from week four onwards. The discharge placement bias effects have more cohort-level variability after the first week. In Fig.~\ref{fig:select}, we zoom in on the effects for the cohorts that benefit the most from the discharge placement interventions.
 A key advantage of this form of modeling against even the computationally interpretable ReLU-nets is the ability to perform mesoscopic cohort-level inference and interpretation.
 Cohort-level information facilitates making a model actionable since actions can be applied to subgroups all at once.

\subsection{Understanding the correlations implied by the decomposition}

Let 
\begin{equation}
    \theta^{(\boldsymbol\kappa)} = \sum_{o=0}^O \theta_o^{(\boldsymbol\kappa)},
\end{equation}
where $\mathbb{E}(\theta_0^{(\boldsymbol\kappa)})=\mu_o^{(\boldsymbol\kappa)},$ $\forall o\leq O.$
Suppose that for two multi-indices $\boldsymbol\kappa_1, \boldsymbol\kappa_2$, that

\begin{itemize}
\item $\theta_o^{(\boldsymbol\kappa_1)} = \theta_o^{(\boldsymbol\kappa_2)}$, $\forall o\leq U < O$
\item $\textrm{Cov}\left(\theta_j^{(\boldsymbol\kappa_1)} ,\theta_k^{(\boldsymbol\kappa_2)}  \right) =0$ if $j\neq k$
\item $\textrm{Var}(\theta_o^{\boldsymbol\kappa}) = \sigma_o^2$
\end{itemize}
Then,  
\begin{align}
    \lefteqn{\textrm{Cov}(\theta^{(\boldsymbol\kappa_1)}, \theta^{(\boldsymbol\kappa_2)})} \nonumber \\
    &\quad= \sum_j\sum_k\textrm{Cov}\left(\theta_j^{(\boldsymbol\kappa_1)} ,\theta_k^{(\boldsymbol\kappa_2)}  \right) \nonumber \\
    &= \left({\small\sum_{j=0}^U\sum_{k=0}^U + \sum_{j=U+1}^O\sum_{k=0}^U+ \sum_{j=0}^U \sum_{k=U+1}^O} \right){\small \textrm{Cov}\left(\theta_j^{(\boldsymbol\kappa_1)} ,\theta_k^{(\boldsymbol\kappa_2)}  \right)} \nonumber \\
    &= \sum_{o=0}^U \sigma^2_o + \sum_{o=U+1}^O \textrm{Cov}\left(\theta_o^{(\boldsymbol\kappa_1)} ,\theta_o^{(\boldsymbol\kappa_2)}  \right).
\end{align}

So,
\begin{align}
    \rho(\theta^{(\boldsymbol\kappa_1)}, \theta^{(\boldsymbol\kappa_2)}) &= \displaystyle\frac{\sum_{o=0}^U \sigma^2_o + \sum_{o=U+1}^O \textrm{Cov}\left(\theta_o^{(\boldsymbol\kappa_1)} ,\theta_o^{(\boldsymbol\kappa_2)}  \right)}{\sum_{o=0}^O \sigma^2_o} \nonumber \\
    &=  \frac{\sum_{o=0}^U \sigma_o^2 + \sum_{o=U+1}^O \rho_o \sigma_o^2}{\sum_{o=0}^O \sigma_o^2}.
\end{align} 

\end{document}